\documentclass[
showpacs,floatfix, aps,prl,amsmath,amssymb,
twocolumn,
groupaddress
]{revtex4}

\def\Xint#1{\mathchoice
   {\XXint\displaystyle\textstyle{#1}}
   {\XXint\textstyle\scriptstyle{#1}}
   {\XXint\scriptstyle\scriptscriptstyle{#1}}
   {\XXint\scriptscriptstyle\scriptscriptstyle{#1}}
   \!\int}
\def\XXint#1#2#3{{\setbox0=\hbox{$#1{#2#3}{\int}$}
     \vcenter{\hbox{$#2#3$}}\kern-.5\wd0}}

\def\dashint{\Xint-}

\usepackage{graphicx}
\usepackage{feynmp}

\renewcommand{\r}{{\boldsymbol{r}}}
\renewcommand{\v}{{\boldsymbol{v}}}
\newcommand{\nnabla}{{\boldsymbol{\nabla}}}
\newcommand{\p}{{\boldsymbol{p}}}
\newcommand{\x}{{\boldsymbol{x}}}
\newcommand{\X}{{\boldsymbol{X}}}

\newcommand{\be}{\begin{equation}}
\newcommand{\ee}{\end{equation}}
\newcommand{\bea}{\begin{eqnarray}}
\newcommand{\eea}{\end{eqnarray}}
\newcommand{\bs}{\begin{split}}
\newcommand{\bes}{\begin{equation}\begin{split}}
\newcommand{\ees}{\end{split} \end{equation}}
\newcommand{\es}{\end{split}}

\newcommand{\lpr}{\left(}
\newcommand{\rpr}{\right)}
\newcommand{\lbr}{\left[}
\newcommand{\rbr}{\right]}

\newcommand{\D}{\mathcal{D}}
\newcommand{\E}{\mathcal{E}}
\newcommand{\G}{\mathcal{G}}

\newcommand{\Tr}{\mathrm{Tr}}

\newcommand{\q}{\boldsymbol{q}}
\renewcommand{\S}{\boldsymbol{S}}
\newcommand{\bsigma}{\boldsymbol{\sigma}}

\newcommand{\req}[1]{Eq.~(\ref{#1})}
\newcommand{\reqs}[1]{Eqs.~(\ref{#1})}
\newcommand{\myref}[1]{(\ref{#1})}
\newcommand{\rref}[1]{(\ref{#1})}

\begin{document}

\title{Negative Echo in the 
Density Evolution of Ultracold Fermionic Gases 
}

\author{F.~Fumarola, Y.~Ahmadian, I.L.~Aleiner and B.L.~Altshuler}
\affiliation{Physics Department, Columbia University, New York, NY 10027}

\pacs{03.75.Ss
,05.30.Fk
,03.75.Kk
}
\begin{abstract}

We predict a nonequilibrium critical phenomenon in the space-time 
density evolution of a fermionic gas 
above the temperature of transition into the superfluid phase. On the BCS side of the BEC-BCS crossover, the
evolution of a localized density disturbance exhibits a negative echo at the point of the initial inhomogeneity. Approaching the BEC side, this effect competes with the slow spreading of the density of  bosonic molecules. However, even here the echo dominates for large enough times.  
This effect may be used as an experimental tool to locate the position of the transition.

\end{abstract}
\date{\today}
\maketitle

Superconducting [or superfluid] fluctuations (SF) have a profound effect on the DC transport and the thermodynamic properties of superconductors, leading to the famous Aslamazov-Larkin and Maki-Thompson (MT) corrections to the conductivity \cite{ALMakiThompson,LarkVarl}.  
Valuable information about the role of SF in the emerging correlations of the many-body wave functions is provided by the real time dynamics. Unfortunately, the latter can hardly be probed in solid state experiments.

On the side of atomic physics, recent studies of fermionic ultracold atomic gases (UCAG) with interparticle interaction that can be fine-tuned by exploiting Fano-Feshbach resonances (FFr)\cite{FanoFeshbach,Ketterle4,Wieman,Timmermans2}, have revealed pronounced effects and found clear signatures of superfluidity \cite{Jin,Grimm,Thomas,Bourdel,Hulet,Ketterle1,PSS,Gurarie} in the regime of the so-called BEC-BCS crossover \cite{reviews}. In contrast to solid state systems,
where it is hard to observe a dynamical response to external perturbations, in trapped UCAG it is
possible to probe the
actual evolution of a disturbance in real time and space. 
This offers, among other things, the exciting possibility of
discovering novel manifestations of the peculiar physics of SF.

In this Letter, we analyze the nonequilibrium space and time
evolution of density in the BEC-BCS crossover,
and find that even above the critical temperature SF qualitatively
affect the dynamics. The quantum kinetic equations (QKE) we
derive here, allow us to analyze dynamical effects in the critical region.

\begin{figure}[!tb]\hspace{-.5cm}
\includegraphics[height=1.9in,width=2.9in,angle=0]{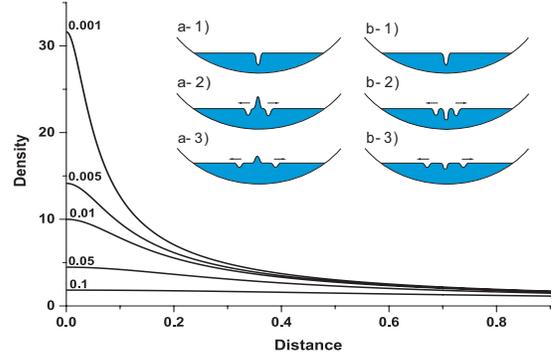}
\vspace{-.4 cm}
\caption{ (Color online) Density profile $\delta \rho_F
(\mathbf{x}, t)$, \req{density},  in arbitrary units 
as a function of
the scaled distance from the origin $\sqrt{\pi b/\gamma(E_1)}m r/t$,
for values of  $\tau$ shown on the curves. 
The inset sketches the total density evolution in regions III-IV (a) and II (b). However, due to the exponential decay of the bosonic dip, even in region II  one always ends up with an algebraically decaying fermionic peak (a-3) for large times.
}
\label{fig1}
\end{figure}

Suppose that originally (at $t=0$) the density is depleted in a small
region inside the system. The time evolution of this
density dip, illustrated by Fig. \ref{fig1}, turns out
to be determined by how close the temperature $T$ is to its critical value $T_c$, and by the position of the system along the crossover.
 In the BCS limit, where $T_c$ is low, we predict
the development of a sharp {\em peak} in the density of fermionic atoms at the origin (see Fig.
\ref{fig1}a). 
The shape of the peak is critical: as $T \to T_c^{+}$
the width shrinks as $\tau^{1/2}$ ($\tau \equiv (T -T_c)/T_c$). As the resonance detuning is further lowered, $T_c$ increases and corrections to the pure BCS
behavior become important: the bosonic molecules depleted initially produce a dip that dominates the fermionic peak for small times and $\tau$'s (see Fig. \ref{fig1}b and region II of Fig.~\ref{regions}). Eventually, however, the bosons decay and the fermionic peak emerges. 
We will find that tuning the system towards the
bosonic side of the crossover reduces the interval of $\tau$'s where the fermionic density peak is
observable. Eventually, in the BEC regime only the bosonic dip remains. To measure these effects, we propose the experiment shown in Fig.~\ref{fig1}a with fermionic UCAG close to the superfluid
transition, following the experimental procedure introduced by
\cite{Andrews}, e.g. by applying the repulsive optical dipole
force of a strongly focused off-resonant laser beam. We stress that the negative echo effect is not limited to the far BCS limit and is observable in the region that is accessible to current experiments.

The profile of the density peak formed by fermions in the BCS side of the crossover, is plotted in  Fig.~\ref{fig1}.
The microscopic mechanism at the root of this phenomenon is the
Andreev reflection of fermionic atoms off the
fluctuations of the superfluid order parameter. 
The wavelength of such fluctuations diverges at
$T \to T_c^{+}$. It is well known \cite{Andreev} that an
electronic excitation, with sub-gap energy, incident on a
superconductor/normal metal interface from the normal side, will
be reflected back as a hole, almost opposite to its original
direction of motion, creating a Cooper pair in the superconductor.
Although the superconducting gap vanishes above the transition
point, close to $T_c$ there appear critically large fluctuating
superfluid islands which transform incident nonequilibrium
particle-like excitations into holes, and vice versa. 
Thus the time evolution shown in Fig. \ref{fig1}a can be understood as
follows. The initial density depletion can be thought of as an
excess of holes, which at $t=0$ start to move radially outwards
from the origin, while the fluctuating condensate reflects them back towards the origin, in the form of atoms
rather than holes. Possible deviation of this
reflection from exact backscattering is determined by the typical
size of a condensate island, which diverges as  $\tau^{-1/2}$. As a result fermions accumulate at the
origin forming a critically sharp density peak, the width of which scales as $\tau^{1/2}$ (see \req{density}).

On the other hand, the density dip of Fig.~\ref{fig1}b, is due to the slow propagation of the bosonic component of the initial nonequilibrium disturbance away from the origin (see \req{bosepeak}). 

The fermionic atoms  interacting via a FFr  
are described by \cite{Eagles, Legget, Schmitt}
\begin{subequations}\label{Model}
\vspace{-.1cm}
\bea 
\hat{H}&=&\int d\r
\left[
\sum_\sigma
\hat{a}^\dagger_\sigma
\epsilon_a\left(\r,-i\hbar\nnabla\right)\hat{a}_\sigma
+
\hat{b}^\dagger\epsilon_b\left(\r,-i\hbar\nnabla\right)\hat{b}
\right]
\nonumber
\\
&+&\frac{g}{2}\int d\r
\sum_{\sigma_1\sigma_2}
\left[\sigma^y_{\sigma_1\sigma_2}\hat{b}
\hat{a}^\dagger_{\sigma_1}\hat{a}^\dagger_{\sigma_2}
+h.c.\right],
\label{model}
\eea
\vspace{-.5cm}
\be
\epsilon_a={\p^2}/({2m})+U_a(\r,t);
\quad \epsilon_b=
\p^2/({4m})+U_b(\r,t),
\label{energies}
\ee
\end{subequations}
where $\hat{a}^\dagger_\sigma(\r),[\hat{a}_\sigma(\r)]$
create [annihilate] fermionic atoms with mass $m$ and spin $\sigma=\pm 1/2$, while
$\hat{b}^\dagger(\r),[\hat{b}(\r)]$ create
[annihilate] the bosonic molecules. $U_{a(b)}$ is the sum of the confining and perturbing potentials
acting on an atom (molecule).

Given the energy of relative motion of two atoms, $\epsilon$, their $s$-channel scattering amplitude can be written as 
\be
f_0(\epsilon)=\hbar\left(m\epsilon\right)^{-1/2}
\gamma(\epsilon)\left[{\epsilon_0-\epsilon-i\gamma(\epsilon)}\right]^{-1},
\label{amplitude}
\ee 
with the resonance width  $\gamma(\epsilon)=\frac{mg^2\sqrt{m\epsilon}}{4\pi\hbar^3}$, and physical position $\epsilon_0$.
 Physical results expressed in terms of $\epsilon_0$ and $\gamma$ are well defined \cite{scattering}.

QKE for the normal phase are derived (see {\em e.g.} Ref.~\cite{LifPit})
by finding a semiclassical approximation for the time ordered Green functions.
We combine the latter into $2\times 2$ matrices,
\be
\hat{\G}=
\begin{pmatrix}
\hat{G}^K & \hat{G}^R
\\
\hat{G}^A & 0
\end{pmatrix};
\quad
\hat{\D}=
\begin{pmatrix}
{D}^K & {D}^R
\\
{D}^A
 & 0
\end{pmatrix},
\label{GF}
\ee
and $\hat{G}^{K/R/A}$ are $2\times 2$ matrices in the spin space.

We limit ourselves to the condition of narrow resonance,
(see Ref.~\cite{strecker} for the experimental realization)
\be
\gamma(E) \ll E,
\ E=\mathrm{max}\left(T,\epsilon_F\right), \
2m\epsilon_F\equiv\left[3\pi^2\hbar^3 \rho_a\right]^{\frac{2}{3}}\!,
\label{narrow}
\ee
where $\rho_a$ is the atomic density in the absence of the resonance,
and $T$ is the temperature \cite{T}. 
For $\tau\gg Gi$, where $Gi\ll 1$ is the Ginzburg parameter \cite{ginzburg}, condition \rref{narrow} justifies the one loop approximation 
\vspace{-.6cm}
\begin{subequations}
\label{dys}
\bea
&&\hspace*{-.7cm}
\raisebox{-2pt}{
\setlength{\unitlength}{0.025in}
\begin{fmffile}{babam1}
\begin{fmfgraph*}(25,15)
\fmfleft{i1,d1}
\fmfright{o1,d2}
\fmf{plain_arrow,label=$\hat{\G}$,width=3}{o1,i1}
\end{fmfgraph*}
\end{fmffile}}
=\hspace{-.3cm}
\raisebox{-2pt}{
\setlength{\unitlength}{0.025in}
\begin{fmffile}{babam2}
\begin{fmfgraph*}(26,15)
\fmfleft{i1,d1}
\fmfright{o1,d2}
\fmf{plain_arrow,label=$\hat{\G_0}$,width=1}{o1,i1}
\end{fmfgraph*}
\end{fmffile}}\hspace{-.2cm}
+ \hspace{-.6cm}
\raisebox{-2pt}{
\setlength{\unitlength}{0.025in}
\begin{fmffile}{babam5}
\begin{fmfgraph}(55,19)
\fmfleft{i1,d1}
\fmfright{o1,d2}
\fmf{plain_arrow,width=1,tension=0.7}{i2,i1}
\fmf{plain_arrow,width=3,tension=0.5}{i2,i3}
\fmf{plain_arrow,width=3}{o1,i3}
\fmffreeze
\fmftop{t3,t4}
\fmf{phantom}{i20,t3}
\fmf{phantom,tension=15}{i2,i20}
\fmf{phantom}{i30,t4}
\fmf{phantom,tension=15}{i3,i30}
\fmfv{decoration.shape=triangle,decoration.filled=0.5,decoration.size=11}{i20,i30}
\fmftop{t1,t2}
\fmf{phantom,tension=1.8}{t1,v4}
\fmf{phantom,tension=1.8}{t2,v5}
\fmf{photon,right=0.5,width=3}{i3,v5}
\fmf{photon,right=0.5,width=3}{v4,i2}
\fmf{plain_arrow,right=0.1,tension=5,width=3}{v5,v4}
\end{fmfgraph}
\end{fmffile}
}
\label{dysonF}
\\
&&\hspace*{-.7cm}
\raisebox{2pt}{
\setlength{\unitlength}{0.025in}
\begin{fmffile}{babat1}
\begin{fmfgraph*}(27,15)
\fmfleft{i1,d1}
\fmfright{o1,d2}
\fmf{photon,width=3}{o1,i3}
\fmf{plain_arrow,label.side=right,label=$\hat{\D}$,width=3,tension=1.1}{i3,i4}
\fmf{photon,width=3}{i1,i4}
\end{fmfgraph*}
\end{fmffile}}
=\hspace{-.5cm}
\raisebox{2pt}{
\setlength{\unitlength}{0.025in}
\begin{fmffile}{babat2}
\begin{fmfgraph*}(27,15)
\fmfleft{i1,d1}
\fmfright{o1,d2}
\fmf{photon,width=1}{o1,i3}
\fmf{plain_arrow,label.side=right,label=$\hat{\D}_0$,width=1,tension=1.1}{i3,i4}
\fmf{photon,width=1}{i1,i4}
\end{fmfgraph*}
\end{fmffile}}
\hspace{-.1cm}
+ \hspace{-.5cm}
\raisebox{2pt}{
\setlength{\unitlength}{0.025in}
\begin{fmffile}{babat6}
\begin{fmfgraph}(57,15)
\fmfleft{i1,d1}
\fmfright{o1,d2}
\fmf{plain_arrow,width=1,tension=1.9}{i2,i1}
\fmf{photon,width=1}{i2,v1}
\fmf{phantom,width=1,tension=0.3}{v1,v2}
\fmf{photon,width=3,tension=1}{i6,v2}
\fmf{plain_arrow,width=3,tension=1.3}{o1,i6}
\fmfv{decoration.shape=triangle,decoration.filled=0.5,decoration.size=11,decoration.angle=90}{v1}
\fmfv{decoration.shape=triangle,decoration.filled=0.5,decoration.size=11,decoration.angle=-90}{v2}
\fmffreeze
\fmf{fermion,left=0.5,width=3}{v2,v1}
\fmf{fermion,right=0.5,width=3}{v2,v1}
\end{fmfgraph}
\end{fmffile}}
\label{dysonB}
\\
&&\nonumber
\\
&&\hspace*{-0.3cm}
\raisebox{-6pt}{
\setlength{\unitlength}{0.025in}
\begin{fmffile}{babar1}
\begin{fmfgraph*}(17,12)
\fmfv{decoration.shape=triangle,decoration.filled=0.5,decoration.size=11,decoration.angle=90}{v}
\fmfleft{i1}
\fmfright{o1,o2}
\fmf{photon}{v,i1}
\fmf{plain}{o1,v}
\fmf{plain}{o2,v}
\fmfv{label.dist=-15pt,label=$\raisebox{12pt}{\scriptsize{$k$}}$}{v}
\fmflabel{\scriptsize{$ l,\sigma_1$}}{o1}
\fmflabel{\scriptsize{$ m,\sigma_2$}}{o2}
\end{fmfgraph*}
\end{fmffile}
}
\ \ \ \  =\frac{-ig\sigma^y_{\sigma_1\sigma_2}d_{klm}}{\sqrt{2}},
\label{dysonC}
\eea
\end{subequations}

\noindent
for the derivation of the QKE.
Here the structure of the vertices in the Keldysh space \rref{GF} is
defined by $d_{klm}=\delta_{k2}\delta_{lm}
+\delta_{l2}\delta_{km}
+\delta_{m2}\delta_{kl}
-
2\delta_{l2}\delta_{k2}\delta_{m2}$.
As usual \cite{LifPit}, the derivation of the QKE
requires obtaining the equation
for the  Wigner transforms  for the Green function ($\r_{1,2}\equiv \r\pm \delta\r/2,
\ t_{1,2}\equiv t\pm \delta t/2$):
\be
\begin{split}
\begin{bmatrix}
\hat{\G}(\r_1,\! t_1;\r_2,\! t_2)\\
\hat{\D}(\r_1,\! t_1;\r_2,\! t_2)
\end{bmatrix}\!
=\!\int\!\frac{d\p\, d\epsilon}
{(2\pi\hbar)^4}\!\ e^{\frac{i(\p\delta\r-\epsilon \delta t)}{\hbar}}
\begin{bmatrix}
\hat{\G}\left(\epsilon,\X\right)
\\
\hat{\D}\left(\epsilon,\X\right)
\end{bmatrix},
\end{split}
\label{Wigner}
\ee
and  keeping leading terms in $\hbar$ expansion.
[Hereinafter, we use the notation $\X\equiv (\p,\x),
\ \x\equiv(\r,t)$.]

We parametrize the Keldysh components
of matrices \rref{GF} in terms of the distribution functions $\hat{n}(\X),N(\X)$: 
\be
\begin{split}
&\hat{G}^K(\epsilon,\X)=-i\pi\left[\delta\left(\epsilon-\hat{\E}(\X)\right),1-2\hat{n}(\X)\right]_+,
\\
&D^K(\epsilon,\X)=-2i\pi\delta\left(\epsilon-\Omega(\X)\right)
\left[
2 N(\X)+1
\right]
, 
\end{split}
\raisetag{4.3em}
\label{parametrization}
\ee
with $\hat{n}(\X)$ a $2\times 2$ matrix in the spin space.
Hereinafter,  $\left[\cdot,\cdot\right]_+$ means anticommutator.
Using \reqs{dys}--\rref{parametrization},
we derive the QKE (details will be published elsewhere):
\begin{subequations}\label{KQE}
\bea
&&\!\!\!\!\!\!\!\!\!\!{\partial_t\hat{n}}\!+\!\frac{i}{\hbar}\left[\hat{\E},\hat{n}\right]
+ \left\{\hat{\E},\hat{n}\right\}_P\!\! =
\!\!\int\!\frac{d\p_1}{\left(2\pi\hbar\right)^3}
\,\hat{\mathrm St}\left( \p,\p_1;\x\right);\label{fermionico}
\\
&&\!\!\!\!\!\!\!\!\!\!{\partial_t N}
+ \left\{{\Omega},{N}\right\}_P\!\!=-\!\int\!\!\frac{d\p_1}{2\left(2\pi\hbar\right)^3}
\,\Tr\,\hat{\mathrm St}\left(\p-\p_1,\p_1;\x\right),\,\,\label{bosonico}
\raisetag{2em}
\label{keq1}
\eea
\end{subequations}
where $\left[\cdot,\cdot\right]$ stand for the commutator
and the Poisson brackets for arbitrary matrices $\hat{A},\hat{B}$ are defined
as $
2\big\{\hat{A},\hat{B}\big\}_P\!\!\!=
\big[\nnabla_\p\hat{A},\nnabla_\r\hat{B}\big]_+
- \big[\nnabla_\r\hat{A},\nnabla_\p\hat{B}\big]_+
$.  Equation \myref{keq1} is valid when the bosonic spectral width is smaller than temperature, and thus is not valid in the far BCS region (region IV of Fig.~\ref{regions}). However, it can be shown that in this region it suffices to use the equilibrium bosonic distribution (see \req{equilibrium})  in \req{fermionico}, and \req{keq1} is not needed.

In \req{keq1}, the energies of the fermionic and bosonic excitations, $\hat{\E}$ and $\Omega$, 
depend on the momentum and the coordinate through the bare energies \rref{energies}
and also on the distribution functions of the excitations themselves.
Due to a possible non-equilibrium spin population, the spectrum $\hat{\E}$
may posess a spin space structure:
\begin{subequations}
\bea
&&\hat{\E}\left(\X;
\left\{\hat{n},N\right\}\right)=\frac{\p^2}{2m}+ 
U_a(\r,t)
\label{epsilon}
\\
&& -
\frac{g^2}{2}\dashint\!\frac{d\p_1}{\left(2\pi\hbar\right)^3}\!
\overline{
\left[
\hat n(\p_1,\x)+N(\p+\p_1,\x),\frac{1}{\hat{\Delta}(\p,\p_1;\x)}
\right]}_+\!\!,
\nonumber
\eea
where notation
$\hat{\Delta}\equiv\Omega(\p_1+\p_2,\x)-\hat\E(\p_1,\x)-\hat\E(\p_2,\x)$ is introduced
and overline signifies the time reversal operation in the spin space:
$\overline{\hat{A}}\equiv \hat{\sigma}_y\hat{A}^T \hat{\sigma}_y$.
The spectrum of the bosonic excitations acquires a shift
\bea
&&\Omega\left(\X;
\left\{\hat{n}\right\}\right)=\frac{\p^2}{4m}+ 
U_b(\r,t)
\label{Omega}
\\
&& \quad +  g^2\dashint\!\frac{d\p_1}{\left(2\pi\hbar\right)^3}
\left[
\Tr
\frac{1/2-\hat n(\p_1,\x)}{\hat{\Delta}(\p-\p_1,\p_1,\x)}
+
\frac{m}{\p^2_1}\right],
\nonumber
\eea
 which for a sharp fermionic distribution function, diverges logarithmically as in conventional BCS. 
\end{subequations}

Real collision processes defining the irreversible evolution in \req{keq1}
are described by (we omitted $\x$ in the argument of the distribution functions)
{
\setlength\arraycolsep{-1.5pt}
\bea
&&\hat{\mathrm St}\left(\p_1,\p_2,\x\right)
=
\frac{\pi g^2}{\hbar}
\left[
\delta\left(\hat{\Delta}(\p_1,\p_2;\x)\right); \hat{\Upsilon}(\p_1,\p_2;\x)
\right]_+\! ;
\nonumber\\
&&
\hat{\Upsilon}=\!
N(\p_1+\p_2)\left[1-\hat{n}(\p_1)-\!\overline{\hat{n}(\p_2)}\right]
- \hat{n}(\p_1)\overline{\hat{n}(\p_2)}.
\label{stoss}
\eea
The QKE scheme \rref{parametrization}--\rref{stoss}
is justifiable as long as the temporal [spatial] variation of the distribution
functions is smooth on quantum  scale, $\hbar /T$  [$\hbar/{\sqrt{mT}} \mathrm{ max}(1,
\sqrt{\epsilon_F/T}
)$] \cite{T}. Under these conditions, the expressions for the collision
integral and spectral renormalization are local in time and space \cite{magnitude}.

For densities of fermions, $\rho_a(\x)$, bosons $\rho_b(\x)$, and spin, $\S(\x)$, we have
\be
\left[\rho_a, 2 \S, \rho_b\right](\x)
=\!\!\int\!\!\!\frac{d\p}{(2\pi\hbar)^{3}}\!
\left[\Tr\,\hat{n},\Tr\,\bsigma\hat{n},
 N\right](\X) .
\label{densities}
\ee

Quantum Kinetic Equations \rref{KQE} respect the symmetries of model \rref{Model}: conservation of mass, energy, spin rotation invariance, and  when $\nnabla U_{a,b}=0$, Galilean invariance. The latter is implemented by
\be
\begin{split}
\hat{n}(\p,\r) &\to \hat{n}(\p-m\v,\r-\v t);\\
N(\p,\r) &\to 
N(\p-2m\v,\r-\v t);
\\
{\hat\E}(\p,\r) &\to {\hat\E}(\p-m\v,\r-\v t)+\v\cdot\p-{m\v^2}/{2};
\\
{\Omega}(\p,\r) &\to {\Omega}(\p-2m\v,\r-\v t)+\v\cdot\p-m\v^2.
\end{split}
\label{Galilean}
\ee
Furthermore, for $\partial_t U_{a,b}=0$ the equilibrium distributions 
\be
\hat{n}={\mathcal F}\!\left(\frac{\hat{\E}(\X)-\mu-\hat{h}}{T}
\right)\!,
\
N={\mathcal B}\!\left(
\frac{\Omega(\X)-2\mu}{T}
\right)\!,
\label{equilibrium}
\ee
where ${\mathcal F}(x)\!=\!1/(e^x+1);\ {\mathcal B}(x)\!=\!1/(e^x-1)$,
solve \reqs{keq1}, with arbitrary constants, $\mu,T$, and an arbitrary traceless Hermitian $2\times 2$ matrix, $\hat{h}$. The latter describes a possible stationary imbalanced spin population, which will be discussed elsewhere. 
 
Transition temperature, $T_c$, is given by \cite{Randeria, KR, Holland, Ohashi, Kokkelmans}
\be
\epsilon_0-2\mu=\!\!\int_{\frac{-\mu}{2T_c}}^{\infty}\!\! \frac{d x}{\pi}\, \gamma(2\mu+4T_c x)\!\left[\frac{\tanh{x}}{x}-\frac{1}{ x+\frac{\mu}{2T_c}}\right]\! .
\label{Tc}
\ee
In the BCS regime, $\epsilon_0\!-\! 2 \epsilon_F\!\gg\! \gamma(\epsilon_F)$, this yields $T_c\! \propto\! \epsilon_F e^{-\frac{\pi}{2}\frac{\epsilon_0 - 2 \epsilon_F}{\gamma(2\epsilon_F)}}$,
and in the BEC regime, ($\epsilon_0\lesssim0$),  $T_c\approx T_{BEC} \simeq 0.218 \epsilon_F$. The equilibrium 
bosonic spectrum, in the interval $T_c Gi <  \Omega(\q) -2\mu \lesssim T$ \cite{ginzburg}, is  $\Omega(\q)=2\mu + c + b\q^2$, with
\bes
\!\!\!c=\!\frac{2\gamma(E_1)}{\pi}\,\tau ;
\qquad b\!=\!\frac{1}{4m}\!\left[1\!+\!\frac{7\zeta(3)}{6\pi^3}
\frac{\gamma(E_2)E_2}{T_c^2}\right]\!.
\label{smallOmega}
\end{split}\ee
Here, $\zeta(3)\simeq 1.202$ (Riemann $\zeta$-function), and
\vspace{-.1cm}
\be
\begin{split}
& \hspace{-.3cm}\begin{bmatrix} 2 E_1^{1/2}\\ \frac{14\zeta(3)}{\pi^2}E_2^{3/2}
 \end{bmatrix}\!\!=\!
\!\!\int\limits_{-\frac{\mu}{2T_c}}^\infty\!\!\!dx\!
\begin{bmatrix}
\frac{1}{\cosh^2 x}
(2\mu+4 x T_c)^{1/2}
\\ \frac{\tanh x - x \mathrm{sech}^2 x}{ x^3}(2\mu+4 x T_c)^{3/2}
\end{bmatrix}\!\! .
\end{split}
\nonumber
\ee
In the BCS limit $T_c\ll \mu$, $E_{1,2}\to 2\mu \approx 2\epsilon_F$. 
\begin{figure}[!t]
\includegraphics[height=125pt,width=190pt,angle=0]{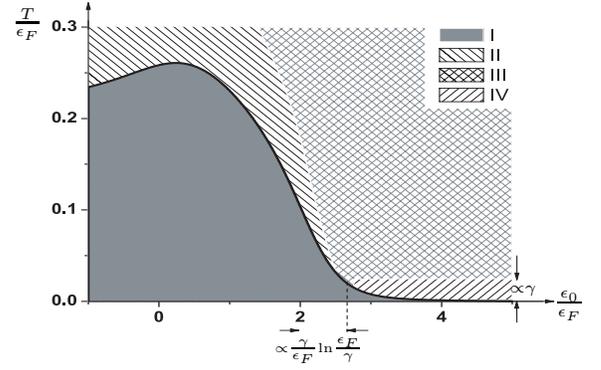}
\put(0,9){$\frac{\epsilon_0}{\epsilon_F}$}
\put(-17,16){$_{\propto \gamma}$}
\put(-106,-3){$ _{_{\propto \frac{\gamma}{\epsilon_F}\!\ln{\!\frac{\epsilon_F}{\gamma}}}}$}
\put(-205,115){$\frac{T}{\epsilon_F}$}
\caption{Relevant regions of parameter space. Region I is the superfluid phase. The QKE \myref{keq1} is valid in II and III. Our results are valid above and close to the critical line (black curve). In III and IV the MT correction,  \req{density}, dominates at all times, whereas in II the bosonic dip,  \req{bosepeak}, dominates for $t<t_*$.}
\label{regions}
\end{figure}\noindent

We now consider the evolution of a small density disturbance as shown
in Fig.~\ref{fig1}. 
A realistic experimental procedure \cite{Andrews} for creating
such a disturbance is by applying a potential $U_a(\mathbf{x})>0$, much narrower than the size of the trap, for a time long enough for the system to equilibrate, with $n(\X, t=0) = \mathcal{F}\left(\frac{\E(p) -\mu+U_a(\x)}{T}\right)$. 
At time $t=0$, the potential is removed abruptly, leaving a small density disturbance.
Next, we linearize the QKE and the inital distributions in the deviation from equilibrium. We solve the linearized fermionic kinetic equation \myref{fermionico} perturbatively in the collision integral, \req{stoss}, but treat the equilibrium decay and spectral renormalization of the bosons non-perturbatively, consistent with our one-loop approximation. 
To zeroth order, the nonequilibrium part of the fermionic distribution evolves as that of free particles: $\delta n^{(0)} (\X,t) = \delta n(\p,\x-\p t/m,t=0)$,  giving rise to an expanding spherical density wave. Similarly for bosons, except for their exponential decay: $ \delta N^{(0)} (\X,t) = \delta N(\boldsymbol{q},\x-2 b \boldsymbol{q}t,t=0)\exp{\lbr -\frac{(c+b \boldsymbol{q}^2)\gamma(2\mu)}{2T_c} t\rbr}$. Deep in the BCS regime (regions III and IV in Fig.~\ref{regions}) the nonequilibrium bosons decay fast and can be neglected. Iterating the fermionic QKE, we find that in this regime, the correction to the 
density close to the origin is dominated by the same terms in the collision integral \myref{stoss} that give rise to the MT correction to the conductivity of conventional superconductors. The latter can also be understood as the effect of Andreev reflection of electrons by superconducting fluctuations. For times $t\gg t_F\equiv \frac{\hbar \mu}{T_c^2 \sqrt{\tau}} $, the density peak normalized by the total number of initially removed fermions  is described (for $r\ll t \sqrt{\frac{2\mu}{m}}$) by 
\be 
\label{density} 
\frac{\delta \rho_F^{(1)}(r,t)}{|\delta \mathcal{N}_F|}\simeq \frac{\gamma(2\mu)\arctan\left(\frac{T_c^2}{\mu}\sqrt{\frac{\pi b}{\gamma(E_1)}} \frac{ m r}{\hbar} \frac{t_F}{t}\right)}{8 
\hbar b\mu r t} .
\ee
In the BCS limit, this result is valid even for a broad resonance, i.e. not restricted by the condition \req{narrow}. 

 In contrast to the exponential decay of any nonequilibrium deviation in the bosonic distribution, this sharp peak in the fermionic density decays only algebraically (one can show that this conclusion is not an artifact of perturbation theory), so that for large times the latter always dominates.  For times smaller than $t^*= \frac{\pi \hbar}{\tau} \frac{T_c}{\gamma(2\mu)\gamma(E_1)}$, and close enough to $T_c$, the increasingly slow moving nonequilbrium  bosons removed by the external potential \cite{ratio} produce a density dip at the origin,
 \be\label{bosepeak}
\! \frac{\delta\rho^{(0)}_B(r,t)}{|\delta \mathcal{N}_B|} \simeq  
 - \frac{v t\sqrt{\tau}}{\pi^2 } \frac{e^{ - \frac{\pi\gamma(2\mu) }{8 b T_c t} \lbr r^2 + \tau v^2 t^2\rbr}}{\lbr  r^2 + \tau v^2 t^2\rbr^2} ,
 \ee
where $v^2 \equiv 8 b \gamma(E_1)/\pi$. This formula is valid for $t\gg t_B\!\equiv \! \hbar/\sqrt{\gamma(E_1) T_c \tau}$ and $r \ll t \sqrt{bT_c}$,
and dominates the fermionic peak \req{density}, reversing the sign of the density deviation at origin (see Fig.~\ref{fig1}b). However, in the BCS regime,  $t^*$ is suppressed, 
and the peak described by formula \myref{density} dominates at all times.
The region where MT correction, \req{density}, dominates is sketched in Fig.~\ref{regions}.

The critical behavior that was found in the response of the system to an external force
may be of valuable use in current experiments on UCAG. It can
provide for instance a new experimental tool to determine the
transition temperature in the BCS side (where the formation of
vortices \cite{Ketterle1} is not a strong enough mechanism), by applying
a small density disturbance on the system and measuring
nondestructively its subsequent evolution.

In summary, we undertook the study of the real time density
evolution of a composite fermion-boson system in the BEC-BCS
crossover, close to the transition point in the normal phase. We
found that in the far BCS regime a localized depletion of density
will evolve into a sharp diverging peak where it was created, as a
manifestation of critical fluctuations of the superfluid
condensate. 
This phenomenon can be exploited in the
experimental study of the crossover in fermionic UCAG's.\\
\indent We are grateful to G.V.~Shlyapnikov  for inspiring conversations that motivated this work, and to G.~Catelani, A.~Andreev, and L.~Glazman for useful remarks.


\begin{thebibliography}{}
\bibitem{ginzburg}{For a narrow resonance, $Gi$ is small throughout the crossover. $Gi\sim \lpr\frac{T_c}{\mu}\rpr^4$ if $T_c\ll \mu$, and $\sim \frac{\gamma^3}{\epsilon_F T_c^2}$ for $\mu<0$.
 }
\bibitem{magnitude}{Typical experiments involve temperatures of order $1 \mu K$, and time resolutions of order $1 ms$, thus satisfying the requirements for the validity of the QKE approach.}
\bibitem{ratio}{The ratio $\frac{\vert \delta \mathcal{N}_B \vert }{\vert \delta \mathcal{N}_F \vert}$  is given by $( \frac{\epsilon_F}{\mu})^{3/2} - 1 $, for $\mu \gg T_c$.}
\end{thebibliography}
\end{document}